\documentclass[11pt]{article}

\RequirePackage[OT1]{fontenc}
\RequirePackage{amsthm,amsmath}
\RequirePackage[numbers]{natbib}
\RequirePackage[colorlinks,citecolor=blue,urlcolor=blue]{hyperref}
\usepackage{setspace}

\RequirePackage{array}
\RequirePackage{amssymb}
\RequirePackage{amsmath}
\RequirePackage{amsthm}
\RequirePackage{amsmath}
\RequirePackage{enumerate}
\RequirePackage{subfigure}
\RequirePackage{graphicx}
\usepackage{amssymb}

\usepackage{amsmath}
\usepackage{cases}
\usepackage{version}
\usepackage{graphicx}
\usepackage{pstricks}

\newtheorem{example}{Example}

\newcommand{\R}{\mathbb R}

\newcommand{\Xb}{\mathbf{X}}
\newcommand{\Yb}{\mathbf{Y}}

\newcommand{\xb}{\ensuremath{\mathbf{x}}}
\newcommand{\yb}{\ensuremath{\mathbf{y}}}

\newcommand{\Ab}{\ensuremath{\mathbf{A}}}

\newcommand{\Ob}{\ensuremath{\mathbf{O}}}

\newcommand{\Sigmab}{{\it{\pmb \Sigma}}}
\newcommand{\deltab}{{\pmb \delta}}

\newcommand{\etab}{{\pmb \eta}}
\newcommand{\betab}{{\pmb \beta}}

\newcommand{\sigmab}{{\pmb \sigma}}
\newcommand{\mub}{{\pmb \mu}}

\newcommand{\zerob}{{\pmb 0}}

\begin{document}

\title{Flexible modelling in statistics: past, present and future}

\author{Christophe Ley \\ \vspace{1cm} \small
ECARES and D\' epartement de Math\' ematique, Universit\'e libre de Bruxelles \vspace{-1cm} \\\small Boulevard du Triomphe, CP210, 1050 Brussels, Belgium \\\small
E-mail address: chrisley@ulb.ac.be}

\date{}

\maketitle

\begin{abstract}
%
%

In times where more and more data become available and where the data exhibit rather complex structures (significant departure from symmetry,  heavy or light tails), flexible modelling has become an essential task for statisticians as well as researchers and practitioners from  domains such as economics, finance or environmental sciences. This is  reflected by the wealth of existing proposals for flexible distributions; well-known examples are Azzalini's skew-normal, Tukey's $g$-and-$h$, mixture  and two-piece distributions, to cite but these. My aim in the present paper is to provide an introduction to this  research field, intended to be useful both for novices and professionals of the domain. After a description of the research stream itself, I will narrate the gripping history of flexible modelling, starring emblematic heroes from the past such as Edgeworth and Pearson, then depict three of the most used flexible families of distributions, and finally provide an outlook on future flexible modelling research by posing challenging open  questions.

\end{abstract}



\normalsize


\section{Introduction.}

\begin{quote}
``Everybody believes in the exponential law of errors : the experimenters,
because they think it can be proved by mathematics; and
the mathematicians, because they believe it has been established by
observation.''\footnote{Original formulation in French: ``Tout le monde y croit cependant, me disait un jour M. Lippmann, car les exp\'erimentateurs s'imaginent que c'est un th\'eor\`eme de math\'ematiques, et les math\'ematiciens, que c'est un fait exp\'erimental.''}

\hfill Lippmann to Poincar\'e, in Poincar\'e (1896), p. 149

\end{quote}\newpage

The ``experimental law of errors'' of course refers to the  normal density 
$$
\frac{1}{\sqrt{2\pi}\sigma}\exp\left(-\frac{(x-\mu)^2}{2\sigma^2}\right)
$$ with location parameter $\mu\in\R$ and scale parameter $\sigma>0$, and the above statement provocatively underlines the often assumed ubiquity of this distribution that is believed to represent the ``normal'' state of data. Indeed, its popularity has numerous sources: the Central Limit Theorem, the strong link to the empirical mean via maximum likelihood characterizations (result due to Gauss 1809, see Duerinckx \emph{et al.} 2014 for a recent detailed account), the maximum entropy characterization (see Cover and Thomas 2006), the numerous well-studied and highly tractable stochastic properties, the bell shape of its curve, the simplicity inherent to the assumption of normality, etc. 
However, more and more empirical evidence has been provided over the years that the assumption of normality is often only a very poor representation of reality, and several data sets cannot be satisfactorily fitted by the normal distribution. 

One nowadays very important domain, where the assumption of normality fails to hold, is finance. Due to the occurrence of extreme events, financial return data have a large amount of probability mass in their tails. Moreover, negative events are usually more extreme than positive events, entailing some form of asymmetry (or skewness) in the data. Both these effects, heavy tails and skewness, which are  considered as stylized facts in the  finance literature, cannot be captured by the normal distribution. I will now describe three further examples of data sets whose behavior goes beyond normality:

\begin{example}\label{exo1}
Stochastic Frontier Analysis (SFA) is concerned with the  specification and estimation of a frontier production function, e.g., for firms. Economic modelling for SFA has been initiated simultaneously by Aigner \emph{et al.}~(1977) and Meeusen and Van den Broeck~(1977) and can be formulated as follows:
\begin{equation}\label{SFA}
Y=f(\xb;\betab)+\epsilon,
\end{equation}
where $Y$ is the  observed scalar output, the production frontier $f$ depends on the input $\xb\in\R^k$ and some  parameter $\betab\in\R^p$ to be estimated, and $\epsilon$ is the error term. This term itself can be expressed as 
\begin{equation}\label{SFA2}
\epsilon=V- U
\end{equation}
 where $V$ is a random shock, assumed to be symmetric, and $U$, independent of $V$, is the random non-negative technical (in-)efficiency component inherent to each firm. 
 Now, the structure \eqref{SFA2} clearly shows that the composed error term $\epsilon$ cannot follow a  normal distribution, since it is the sum of a symmetric term $(V)$ and a negative term~$(-U)$, leading to skewness in the error term.
\end{example}

\begin{example}\label{exo2}
Since the advent of internet, analysing internet traffic data has become an important field of study. Ramirez-Cobo \emph{et al.}~(2010) have investigated a trace of LAN and WAN traffic recorded in 1989, more precisely, the measured transferred bytes/sec within 3142 consecutive seconds. The log-transformed data are shown in Figure~\ref{Fig0}.

\begin{figure}
\begin{center}
\vspace{6mm}
\begin{minipage}{110mm}
\resizebox*{9cm}{!}{\includegraphics{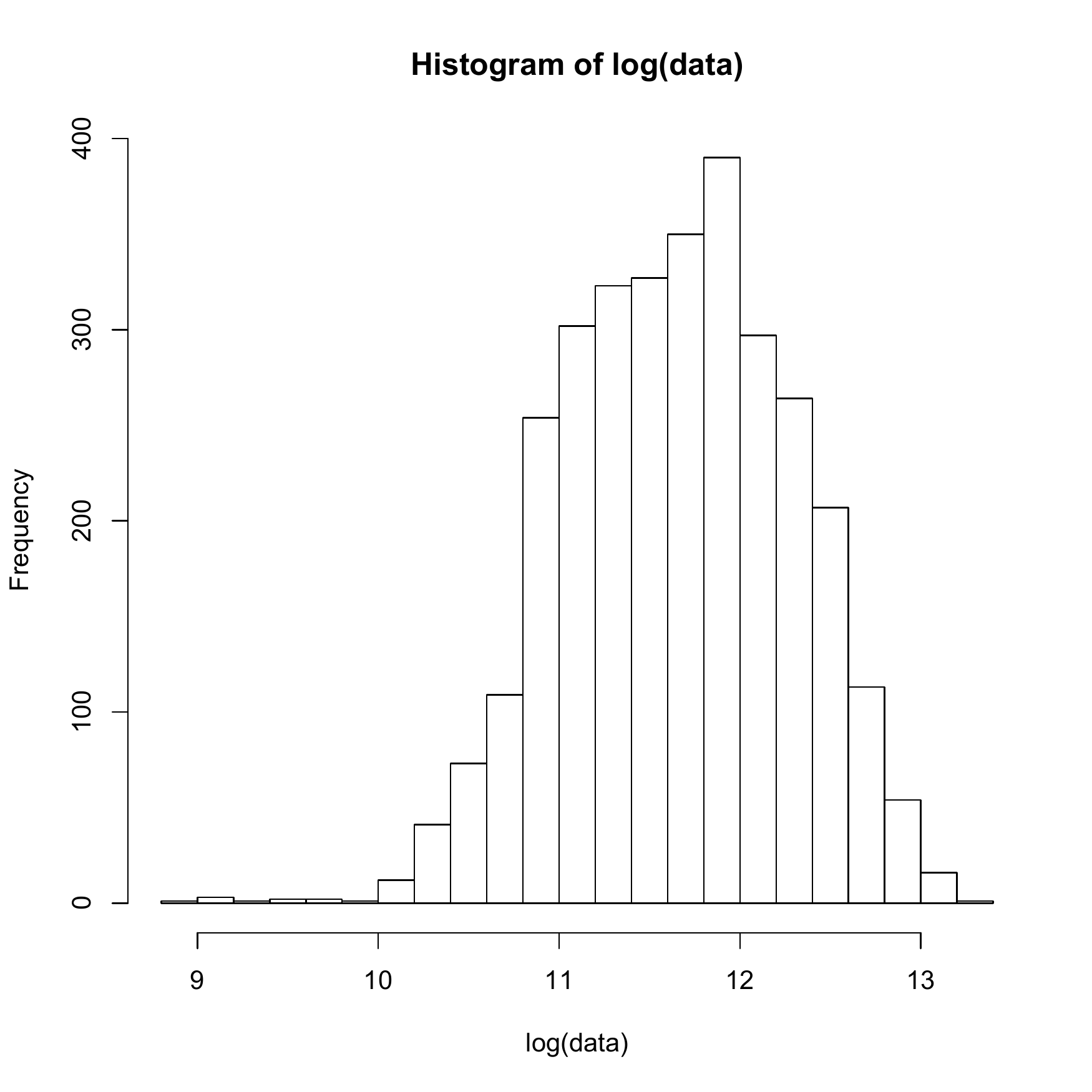}}%
\caption{Histogram of log-scaled internet traffic data (Example~\ref{exo2}).}%
\label{Fig0}
\end{minipage}
\end{center}
\end{figure}

One readily observes that the right tail of the log-scaled internet traffic data is lighter than the  left tail, indicating the need for a model that can incorporate non-normal tail-weight (the right tail appears lighter-than-normal) as well as asymmetry.
\end{example}

\begin{example}\label{exo3}
Given the thread that obesity represents to a human's health, the Body Mass Index (BMI, in ${\rm kg}/{\rm m}^2$), calculated as the ratio between the body weight in kilograms and the squared body height in meters, has become an essential tool to determine overweight and obesity. Lin \emph{et al.}~(2007) have analyzed the BMI for men between the age of 18 and 80 with weights either ranging from 39.5 kg to 70 kg (1069 individuals) or from 95.01 kg to 196.8 kg (1054 individuals). 

Due to the strong correlation between body weight and BMI, it is no surprise that the resulting BMI data are bi-modal. Moreover, given the different ranges of weight, it is also clear that the right tail (high BMI) is heavier than the lower tail (small BMI). These evident observations exclude the normal distribution as a possible fit for BMI data.
\end{example}

These examples, on which I shall come back later on in this paper, further underline the clear need for distributions exhibiting features that the normal does not possess. In other words, the need for \emph{flexible distributions}. Now, one may think that this is an easy task---heavy tails can be obtained by replacing the normal with the Student $t$ distribution, and skewness can be attained by simply mixing two normal  distributions. This is still too much of a simplified vision, as these two parametric models also have their limitations, especially when one has to combine skewness and excess kurtosis in a specific way, as is the case with financial data. As a reaction to such limitations, one may want to consider models with a very large number of parameters in order to ensure as many different shapes as possible. Such a solution in turn suffers from a risk of over-fitting and complexity in calculations. 
%

Thus the ideal solution is a compromise between both requirements: wide range of possible shapes and tractability. This is precisely the objective of the research stream called \emph{flexible modelling}, and the topic of the present paper. The importance of flexible models is further stressed by their successful combination with classical statistical theories and tools such as time series analysis (Hansen~1994), space-state models (Naveau \emph{et al.}~2005), random fields (Allard and Naveau~2007), regression models (Azzalini and Genton~2008), Bayesian statistics (Rubio and Steel~2014) or hypothesis testing (see Section~\ref{sec:fut}), to cite but a few. 

The paper is organized as follows. In Section~\ref{sec:past}, I will shortly narrate the beginnings of flexible modelling, and then describe three of the nowadays most popular flexible families in Section~\ref{sec:pres}. After this review of past and present, I will finish the paper with an outlook on the future of flexible modelling in Section~\ref{sec:fut} by posing (and discussing) six open questions and problems in this field.

\section{A brief historic account: flexible modelling before the 1980s}\label{sec:past}

The systematic quest for non-normal distributions has been initiated at the end of the $19^{\rm th}$ century, when data sets exhibiting marked departures from normality  have been collected, \emph{inter alia} by the famous Belgian scientist Adolphe Quetelet. Due to the pre-dominance and well-understood properties of the ``normal theory'', the first reflex at that time was to transform the data so as to make the resulting transformed data follow a normal distribution, and then apply the usual procedures. For instance,  Galton used in 1979 the logarithmic transformation, which resulted in the nowadays well-known log-normal distribution. It was Francis Ysidro Edgeworth who, in 1898, formally developed this concept, which he coined
``Method of Translation'' instead of transformation. While Edgeworth's transformations were restricted to polynomial functions, the subsequent proposals emerging over the course of the $20^{\rm th}$ were more diverse, the most notable contributions being the Johnson (1949) families and Tukey's $g$-and-$h$ distributions from 1977, both of which will be described in the next section. Still nowadays,  transformations are mainly applied to the normal distribution, for the reasons mentioned above, albeit the Student $t$ and logistic have been shown to be good candidates, too.

The paper Edgeworth (1898) was not Edgeworth's first publication on the subject. Already in the 1880's was he interested in fitting non-normal data, and Edgeworth~(1886) is commonly considered as the first systematic attempt to fitting asymmetric distributions to asymmetric frequency data.  Short time after Edgeworth's publication, another heavy-weight of statistics, namely Karl Pearson, entered the scene of flexible modelling. His interest was raised by the zoologist Walter Weldon who collected data on 23 characteristics of 1000 female crabs during his holidays in Malta and the Bay of Naples and who noticed that one of these characteristics did not possess normal-like behavior. Pearson's first attempts resulted in a mixture of two normal distributions (hereby laying the foundations of mixture distributions) as well as in a ``generalized form of the normal curve of an asymmetrical character'' (this was actually a precursor of the Gamma distribution), both published in a letter to \emph{Nature} (Pearson 1893). This publication triggered a sort of competition between Edgeworth and Pearson. For instance, Edgeworth  drew Pearson's attention to the fact that the Gamma-like curve had previously been derived by Erastus De Forest in De Forest (1882, 1883), a fact that Pearson acknowledged in a further letter to \emph{Nature} (Pearson~1895a). 

Pearson's next attempt on skew frequency curves was even more groundbreaking: in Pearson (1895b), he defined several probability distributions as solutions to a particular differential equation. This seminal idea formed the basis of the thoroughly studied and still in modern days used \emph{Pearson family of distributions} (see Johnson \emph{et al.}~1994 for a detailed description of that family), which he further developed in Pearson (1901) and Pearson (1916). Among the 1895 distributions figures the so-called Pearson Type IV distribution, which contains as special case the Student $t$ distribution; Pearson thus outdates William Sealy Gosset's 1908 paper written under the pseudonym ``Student''. Edgeworth's reaction to Pearson's works was the already mentioned Edgeworth (1898) paper; the latter was further backed up by Kapteyn (1903), the second father of the transformation approach. On top of the Pearson-Edgeworth rivalry, a heated exchange started between Pearson and Kapteyn, each promoting his own family of skew curves and criticizing the other. The Pearson-Kapteyn dispute is reported in Stamhuis and Seneta (2009), while the competition between Edgeworth and Pearson and their constant correspondence has been mostly reprinted in Stigler (1978); see also the excellent Stigler (1986) for further historical details about that fruitful period. 

Exactly at that same time appeared the \emph{Kollectivmasslehre} by Carl Gustav Fechner (Fechner 1897), the founder of psychophysics, the science studying the link between psychological sensations and physical stimulus. Fechner died in 1887, and it was Gottlob Friedrich Lipps who took on the heavy task of completing his manuscript, which explains the posthumous publication in 1897. In that work, Fechner proposed a skew curve by binding together two halves of normal curves, each having a different scaling. Fechner thus was the father of what we now call \emph{two-piece distributions}. However, again Pearson vividly opposed himself to Fechner's idea, both from a statistical as well as historical viewpoint. Indeed, he erroneously believed that De Vries (1894) had made the same proposal. On statistical grounds, he argued that, compared to his family of curves, Fechner's was not general enough. This fierce opposition by Pearson\footnote{Quite remarkably, Pearson attacked Fechner's curves in a paper from 1905 where he actually reacted on the work Ranke and Greiner (1904), two anthropologists who claimed that, for their domain, only the normal distribution mattered, and hence disqualified both Pearson's and Fechner's work. Pearson published his reaction in his own journal Biometrika, hereby criticizing Fechner, although both were on the same side w.r.t. Ranke and Greiner. Other remarkable fact: this paper, Pearson (1905), is most well-known as Pearson there introduced the terminology mesokurtic, leptokurtic and platykurtic.} implied that Fechner's work was overlooked for a long time and eventually fell into oblivion. As a consequence, the  wheel has  been re-invented under the names ``joined half-Gaussian'', ``three-parameter two-piece normal'' and ``binormal'' distribution; see Sections 3 and 5 of Wallis (2014) for details about these re-discoveries (including a paper by Edgeworth) and many more details about two-piece distributions. It is to be noted that Fechner's contribution has been re-brought to the statistical community mainly through Hald~(1998).

At the beginning of the $20^{\rm th}$ century, a young Italian scholar, Fernando de Helguero, adopted a completely different viewpoint on \emph{abnormal curves} as he called them. In two fundamental papers, de Helguero (1909a, b), he set up his own program to deal with non-normal data. He therein criticized Edgeworth's and Pearson's works by noting that, although clearly more valuable than the normal since they are broader, their proposals suffer from a major drawback: their constructions do not allow to understand which mechanism might have actually generated the data. To quote de Helguero on their non-normal curves: ``are defective in my view because they only limit themselves to tell us that the infinitesimal elementary causes of variation are interdependent''. His own view is that non-normal behavior only occurs subject to external perturbations, and abnormality is the upshot of some selection mechanism.  The ensuing abnormal curve he derived is the pre-cursor of the celebrated \emph{skew-normal distribution} which, re-appearing under several guises in e.g. Birnbaum (1950), O'Hagan and Leonard (1976) or Aigner \emph{et al.}~(1977), has come to fame thanks to the seminal paper Azzalini~(1985). See the review paper Azzalini (2005) for a more detailed account on these rediscoveries of the skew-normal, and  Azzalini and Regoli (2012a) for a historic analysis of de Helguero's work.

The Azzalini~(1985) paper had a phenomenal impact on  flexible modelling, as it is the starting point for a systematic treatment thereof with growing activity till today. This is why I decide to stop the historical account here\footnote{There would obviously be several further developments to relate such as, for example, the history of  copulas; this would however shift away the focus of the present paper.}; for interested readers, besides the review papers I already cited above, I recommend the Kotz and Vicari (2005) survey paper, the (short) Section 3 of Pewsey (2014) as well as the introductions of the recent PhD theses of Juan Francisco Rosco from the University of Extremadura (Rosco~2012) and of Francisco Javier Rubio from the University of Warwick (Rubio~2013).

\section{Description of the main  families of flexible distributions}\label{sec:pres}

If we credit a distribution with flexibility as soon as its shape significantly diverges from that of the normal distribution, then of course the task of this section is near-impossible given the infinity of non-normal distributions. The historical developments of the previous section teach us, in my opinion, an important lesson to which I shall stick throughout the rest of the paper: flexible models ought to be distributions that, besides the usual location and scale/scatter parameters,  possess either a skewness or a kurtosis parameter, or, optimally, both. Now this still leaves us with a plethora of distributions, ranging from the Pearson family to the hyperbolic  and   the Tukey lambda distribution  by passing  across the $\alpha$-stable and generalized extreme value families. 

I here ``restrict'' my attention to flexible modelling understood as modifying a given base density $f$, symmetric about the origin\footnote{In higher dimensions, symmetry of $\Xb$ with values in $\R^k$ shall mean, according to the situation, spherical symmetry ($\Xb\stackrel{d}{=}\Ob\Xb$ for all orthogonal $k\times k$ matrices $\Ob$), elliptical symmetry ($\Xb\stackrel{d}{=}\Ab \Yb$ where $\Yb$ is spherically symmetric and $\Ab$ is a full-rank $k\times k$ matrix) and central symmetry ($\Xb\stackrel{d}{=}-\Xb$).}.  Most constructions  nowadays follow this seemingly natural pattern. The key advantage of modifying a symmetric density $f$ is that in doing so it is possible to retain
some of the properties of $f$, which are often well known. If two  parameters are added, $f$ is mostly normal; in case a single skewness parameter is added, $f$ is mostly the Student $t$ density. To avoid any misunderstandings, I provide the latter density in the $k$-variate case:
$$ 
t_{\mub,\Sigmab,\nu}(\xb;k):=\frac{\Gamma \left(\frac{\nu+k}{2}\right)}{(\pi \nu)^{k/2} \,\Gamma \left(\frac{\nu}{2}\right)} 
|\pmb{\Sigma}|^{-1/2}\left(1+\|\pmb{\Sigma}^{-1/2}(\pmb{x}-\pmb{\mu})\|^2/\nu \right)^{-\frac{\nu+k}{2}}
$$
with location parameter $\pmb{\mu} \in \mathbb{R}^k$, scatter parameter  $\pmb{\Sigma} \in S_k$, the class of symmetric and positive definite $k \times k$ matrices, and tail-weight parameter $\nu \in \mathbb{R}_0^+$, and where the Gamma function is given by $\Gamma(z)=\int_{0}^\infty \exp(-t)t^{z-1}\, \mathrm{d}t$. Throughout this section, $\mub$ and $\sigma/\Sigmab$ are location and scale/scatter parameters, respectively.

In what follows, I will depict  three of the most used families of flexible distributions, namely \emph{Azzalini-type distributions}, \emph{transformation-approach distributions} and \emph{two-piece or scale-transformed distributions}; the order of presentation is in line with Jones~(2014a).  Evidently, this enumeration could easily be extended by other very popular flexible modelling families and tools such as finite mixture models (McLachlan and Peel~2000), variance-mean mixtures (Barndorff-Nielsen \emph{et al.}~1982), copulas (Nelsen 2009), power transformations including the Box-Cox transformation (Box and Cox~1964), order-statistics-based distributions (Jones~2004), the very general probability integral transformations of Ferreira and Steel~(2006) or the classical Pearson system of distributions.  However, such a broad description would be far beyond the scope of the paper and dilute its main focus of providing a concise idea of the flexible modelling research. Also, I here do not consider data on supports other than $\R^k$, although much could be said about data on finite or semi-finite intervals (e.g., the logarithmic and power transformations) or on directional data (see  Section~\ref{sec:fut}). For further general information on flexible distributions, see my encyclopedic paper Ley (2012), whose focus is solely on skew distributions, the contribution Lee \emph{et al.}~(2013) and the excellent discussion paper Jones~(2014a), where four flexible families are compared in terms of their stochastic and statistical properties. To avoid any redundancy, I will not proceed to such a comparison and confine my-self to a pure description, which is intended to differ as much as possible from the aforementioned  references.


\subsection{Family 1: Azzalini-type distributions}

The construction underlying this first family of distributions is purely of a skewing nature, although tails can partly be affected, {see Ferreira and Steel~(2006, Section 3.1). This nature of Family 1 is underpinned by the terminology ``modulating symmetry'' advocated by Adelchi Azzalini in recent years. The modulation can well be seen from~\eqref{SSym} below: the symmetric part is multiplied by a \emph{skewing function}. This pure-skewing construction explains why the skew-$t$ distribution defined  in~\eqref{skewt} is a popular choice in flexible modelling. Speaking of popularity, Family 1 has encountered an enormous success over the past decades, with an incredible number of papers dedicated to it. Reasons herefore are  good stochastic properties, nice generating mechanisms (see the part on de Helguero of the previous section), good fitting properties and, ``bad'' critics being better than no critics, some unfortunate inferential problems. This cocktail of attributes has helped to forge the fame of what one may call Azzalini-type distributions. It is therefore also not surprising that there is nearly no ambiguity when speaking about \emph{the} skew-normal distribution: it is Azzalini's from 1985 (his systematic treatment and study of the law  makes him the father of the skew-normal, despite previous similar proposals as mentioned in Section~\ref{sec:past}).

At the core of this first family   figures evidently the (scalar) skew-normal  density 
$$
2\sigma^{-1}\phi\left(\frac{x-\mu}{\sigma}\right)\Phi\left(\delta\left(\frac{x-\mu}{\sigma}\right)\right)
$$
with $\delta\in\R$ the skewness parameter and where $\phi$ and $\Phi$ respectively denote the standard normal density and cumulative distribution function (cdf). The role of $\delta$ as skewness parameter becomes most apparent by noting that only at $\delta=0$ the skew-normal density is symmetric, namely on retrieves the normal density. The multivariate skew-normal has been proposed some ten years later in Azzalini and Dalle Valle~(1996) and further studied in Azzalini and Capitanio~(1999). Subsequent generalizations include skew-elliptical (skewing an elliptically symmetric density), generalized skew-elliptical and, most generally, \emph{skew-symmetric} distributions. The latter, proposed by Azzalini and Capitanio~(2003) and Wang \emph{et al.}~(2004), admit densities of the form
\begin{equation}\label{SSym}
2 |\pmb{\Sigma}|^{-1/2} f\left(\Sigmab^{-1/2}(\xb-\mub)\right) \Pi\left(\Sigmab^{-1/2}(\xb-\mub),\deltab\right)
\end{equation}
where $f$  is centrally symmetric and $\Pi:\R^k\times\R^k\rightarrow[0,1]$ is a {skewing function} satisfying $\Pi(\yb,\deltab)+\Pi(-\yb,\deltab)=1$ and $\Pi(\yb,\zerob)=1/2$ for all $\yb,\deltab\in\R^k$ (for diverse choices of $\Pi$ and resulting skew-$f$ distributions, see Hallin and Ley~2014). For a very recent account on these distributions,  I refer to the book Azzalini and Capitanio~(2014).

Besides these general proposals, parametric skew-$f$ distributions have as well emerged, such as the skew-Cauchy, skew-exponential power, skew-logistic and, most prominently, the skew-$t$ distribution (under various forms). I here consider the multivariate skew-$t$  as defined in Azzalini and Capitanio~(2003), with density
\begin{equation}\label{skewt}
2t_{\mub,\Sigmab,\nu}(\xb;k)T_{0,1,\nu+k}\left(\deltab'\sigmab^{-1}(\xb-\mub)\left(\frac{\nu+k}{|| \Sigmab^{-1/2}(\xb-\mub)||^2+\nu}\right)^{1/2};1\right),
\end{equation}
where $\sigmab$ is a $k\times k$ diagonal matrix with diagonal entries $\sigmab_{ii}=\Sigmab^{1/2}_{ii}$, $i=1,\ldots,k$, and $T_{\mu,\sigma, \eta}(\cdot;1)$ is the cdf of the univariate Student density $t_{\mu,\sigma,\eta}(\cdot;1)$, $\eta>0$. This skew-$t$ distribution is often used for modelling purposes, as it incorporates both a skewness and a tail-weight parameter; see Azzalini and Genton~(2008) for an overview of statistical procedures involving the skew-$t$. For the sake of illustration, I present in Figure~\ref{Fig1} various one-dimensional skew-normal and skew-$t$ densities.

\begin{figure}
\begin{center}
\vspace{6mm}
\begin{minipage}{140mm}
\resizebox*{14cm}{!}{\includegraphics{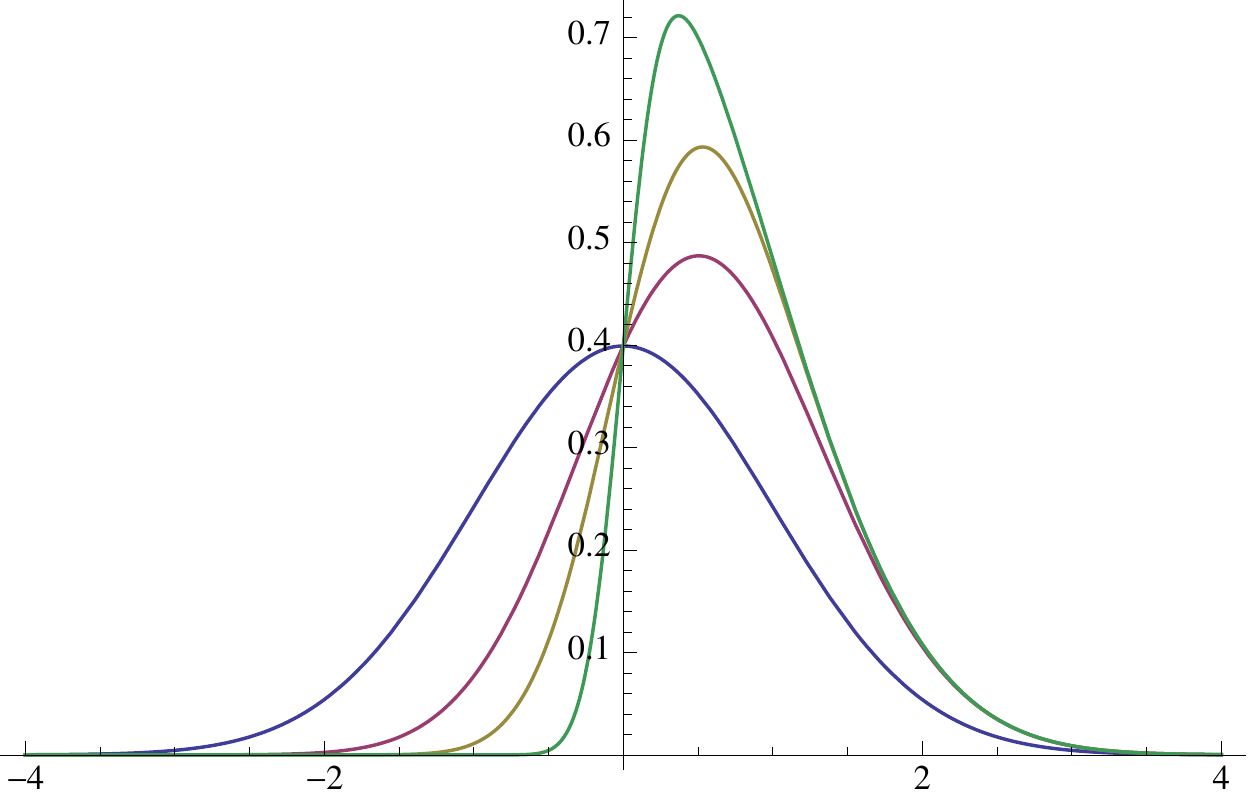}\hspace{1cm}\includegraphics{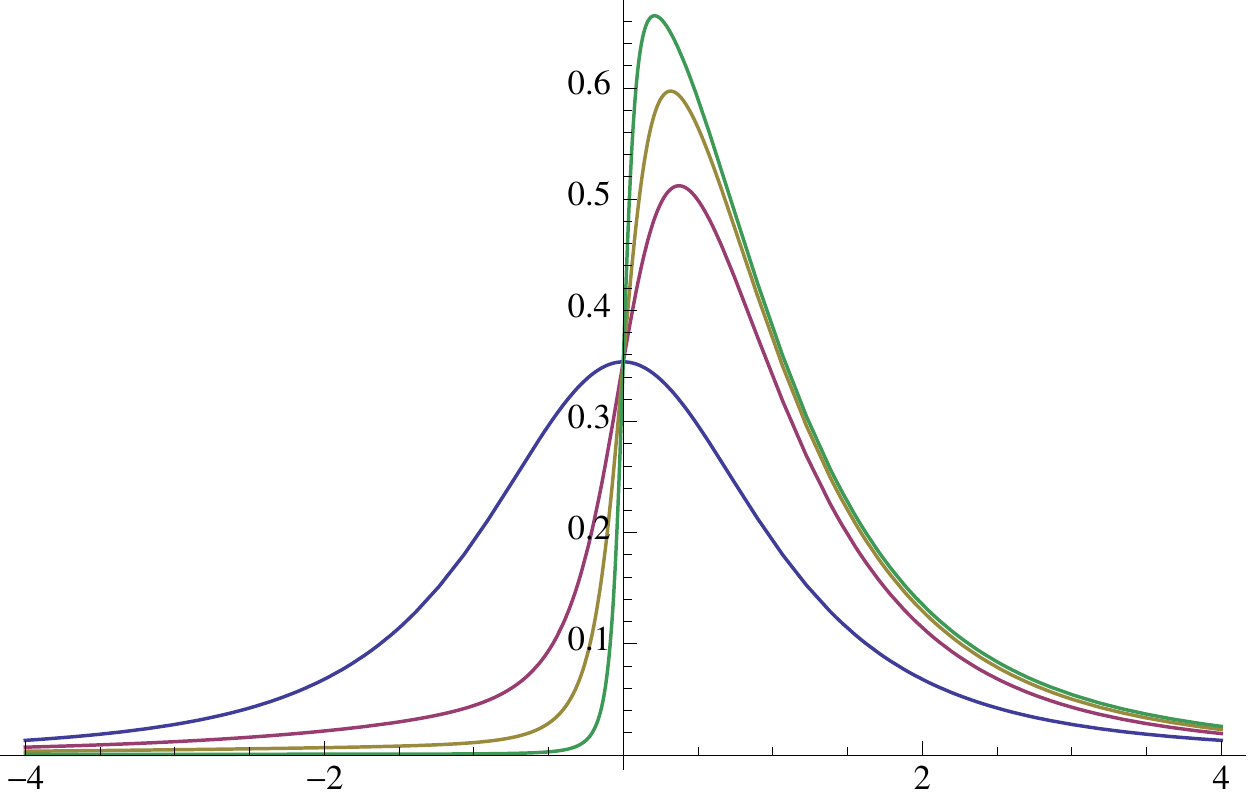}}%
\caption{Plots of the original Azzalini~(1985) skew-normal density $2\phi(x)\Phi(\delta x)$ (left) and of the Azzalini and Capitanio~(2003) skew-$t$ density~\eqref{skewt} with $\mu=0, \sigma=1,\nu=2$ (right),  for $\delta=$0 (blue), 1 (red), 2 (yellow), and~5 (green). }%
\label{Fig1}
\end{minipage}
\end{center}
\end{figure}

The afore-mentioned good properties of skew-symmetric distributions can be found in the references I have given above; see also Azzalini and Regoli~(2012b). In particular, the second part of the monograph Genton~(2004) contains numerous real data situations where skew-symmetric distributions are put to use. Regarding physical generating mechanisms or stochastic representations, I refer the reader to Azzalini~(2005) or to the discussion paper Arnold and Beaver~(2002) where especially hidden truncation and selective sampling are treated. I conclude this section by briefly discussing the peculiar inferential issues related to Family 1.

First, in the vicinity of symmetry, that is, at $\deltab=\zerob$,  certain subfamilies of densities~\eqref{SSym} suffer from a singular Fisher information matrix (w.r.t. $\mub,\Sigmab,\deltab$) due to a (partial) collinearity between location and skewness scores. The most prominent candidate is the skew-normal itself, which is the main reason for the large number of publications on the topic, some trying to explain the phenomena, others to determine all skew-symmetric distributions suffering from that singularity\footnote{Closely related to the Fisher information singularity is another singularity, namely a stationary point at $\deltab=\zerob$ in the profile log-likelihood function for skewness.}, and again others to provide reparameterizations avoiding the phenomena.  
The  question of ``who is infected'' has been solved in the series of papers Ley and Paindaveine (2010a,b) and Hallin and Ley~(2012), while suggested remedies are the centred parameterization of Azzalini~(1985), extended to higher dimensions in Arellano-Valle and Azzalini~(2008), and the Gram-Schmidt-based reparameterization put forward in Hallin and Ley~(2014). In the latter paper, we also classify distributions according to their degrees of singularity (how many orthogonalization and reparameterization steps are necessary to obtain a non-singular matrix), hereby providing intriguing and entertaining insights into the structure of skew-symmetric distributions. Second, for some data sets, the maximum likelihood estimate of $\deltab$ becomes infinite, that is, lies at the frontier of the parameter space. This means that half-normal distributions seem the most appropriate fit to data, although not all data points are positive! This anomaly is well described in Section 6.3 of Azzalini and Capitanio~(1999), and as remedy  Azzalini and Arellano-Valle~(2013) suggest penalized maximum likelihood estimation. In the literature there is no consensus on whether this  is really an issue or simply a natural arte-fact (e.g., Jones~2014a considers the problem as being ``overblown'').

\subsection{Family 2: Transformation-approach distributions}

As mentioned at the beginning of Section~\ref{sec:past}, the initial incentive behind the transformation approach was the pre-dominance of the normal distribution, and, if data were obviously not normally distributed, then one should strive to find a transformation $H$ such that $H(\mbox{data})\approx\mathcal{N}(\mu,\sigma^2)$. The success of the transformation approach has entailed that nowadays the transformations \emph{per se} are interesting, as they lead to very diverse shapes in the resulting density. Indeed, compared to Family 1 and 3, Family 2 is the most flexible, thanks to the freedom of choice in $H$. In the subsequent lines, I will comment on some of the most used such transformations.

In their most general $k$-variate form, transformation-approach densities  built upon some base density $f$ take on the guise
\begin{equation}\label{LP10}
|\Sigmab|^{-1/2}f\left(H_{\deltab,\etab}\left(\Sigmab^{-1/2}(\xb-\mub)\right)\right)\left|DH_{\deltab,\etab}\left(\Sigmab^{-1/2}(\xb-\mub)\right)\right|,
\end{equation}
where $\deltab\in\R^k$ is a skewness parameter and $\etab\in(\R_0^+)^k$ a tail-weight parameter\footnote{Note that both the skewness and tail-weight parameter need not be of  dimension $k$ but could be either larger or smaller.}, and where the transformation $H_{\deltab,\etab}:\R^k\rightarrow\R^k$ is a monotone increasing diffeomorphism with  $|DH_{\deltab,\etab}(\cdot)|$ representing the absolute value of the determinant of the jacobian matrix $DH$. To avoid any ambiguity, I prefer to stress that $H$-transformed densities, let us write them $f_H$, are obtained by transforming $\Xb\sim f$ into $\Yb\sim f_H$ via the (inverse) mapping $H^{-1}_{\deltab,\etab}$ so that $H_{\deltab,\etab}(\Yb)\sim f$. Densities~\eqref{LP10} have been studied in detail in Ley and Paindaveine (2010c) with particular focus on the skewness parameter. For instance, one can easily check that~\eqref{LP10} is symmetric iff $H_{\deltab,\etab}(\cdot)$ is odd. Changes on tail-weight, in terms of largest finite directional moments, can be explicitly calculated via Theorem~3.2 in Ley and Paindaveine~(2010c).  In that same paper, we have identified the minimal requirements on $H_{\deltab,\etab}(\cdot)$ to satisfy the surjectivity property, meaning that every random $k$-vector $\Yb$ can be obtained from any $\Xb$ via one  such transformation. Another popular approach to define multivariate versions of existing univariate transformation-approach distributions consists in transforming each marginal with a one-dimensional transformation $H_{\delta,\eta}$ and introducing some dependence structure by means of a correlation matrix.

The classical Johnson~(1949) paper applies the arcsinh transformation on the normal distribution to obtain the Johnson-$S_U$ distribution, Tukey~(1977) uses $H^{-1}_{\delta,\eta}(x)=\frac{1}{\delta}\left(\exp(\delta x)-1\right)\exp(\eta x^2/2)$ in combination with the normal to define his famous $g$-and-$h$ distributions ($g=\delta$ and $h=\eta$), later extended to higher dimensions in Field and Genton~(2006), Rieck and Nedelman~(1991) replace Johnson's arcsinh transformation with the sinh transformation, whereas Jones and Pewsey~(2009) combine both in their sinh-arcsinh (SAS) transformation $H_{\delta,\eta}(x)={\rm sinh}(\eta\,{\rm arcsinh}(x)+\delta)$ which they also generalize to the multivariate setting. Transformations that affect only tail-weight are, among others, the $K$ transformation $H^{-1}_\eta(x)=x(1+x^2)^\eta$ of Haynes \emph{et al.}~(1997) and the $E$-, $J$-transformations of Fischer and Klein~(2004). For further examples of transformations, see Rosco~(2012) and Rubio~(2013). One may wonder why, in those examples, I switched between writing out $H$ and $H^{-1}$. This is due to the fact that Tukey-type transformations (the $g$-and-$h$ as well as the  $K$-, $E$-, $J$-transformations) do not admit an inverse, hence their density cannot be written out properly and the parameters are often estimated by quantile-based methods. Maximum likelihood estimation, the typical procedure for Family 2, has been examined for Tukey-type distributions in Rayner and MacGillivray~(2002). As an illustration, I provide in Figure~\ref{Fig2} diverse shapes of SAS-normal densities.

\begin{figure}
\begin{center}
\vspace{6mm}
\begin{minipage}{140mm}
\resizebox*{14cm}{!}{\includegraphics{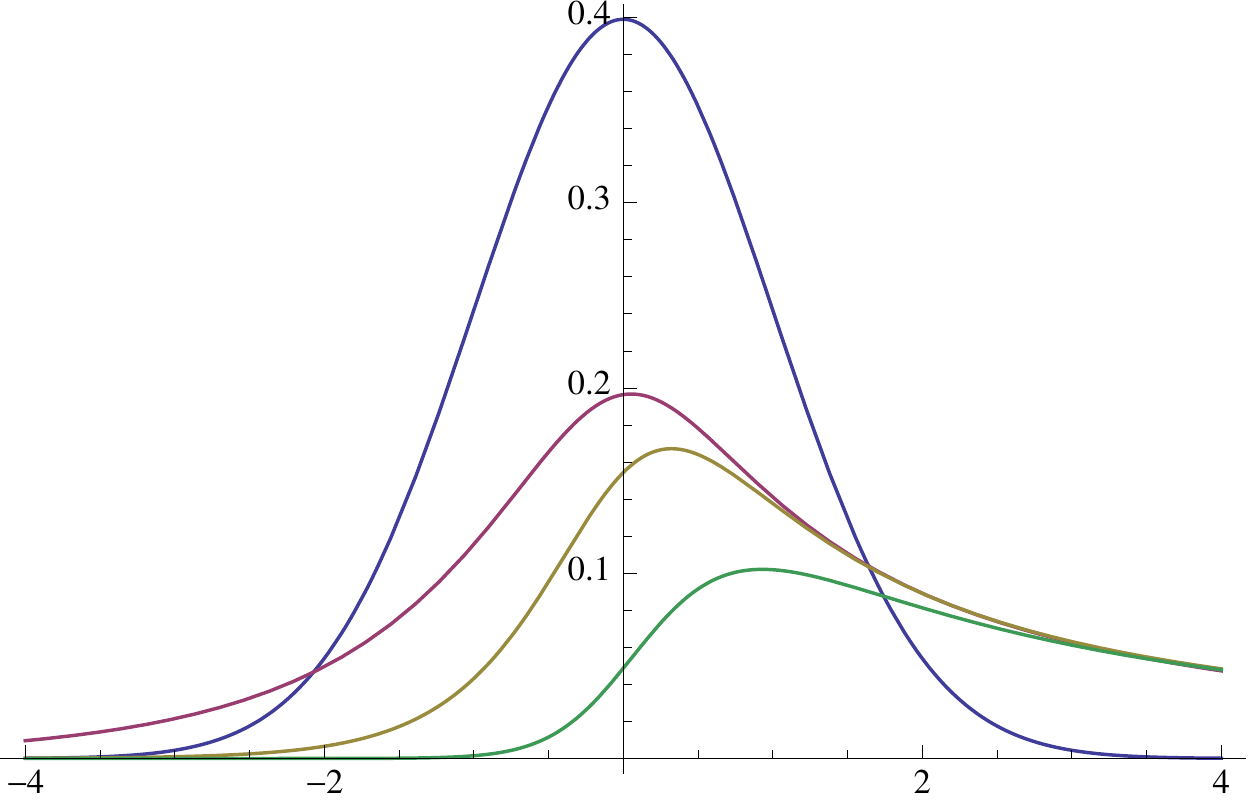}\hspace{1cm}\includegraphics{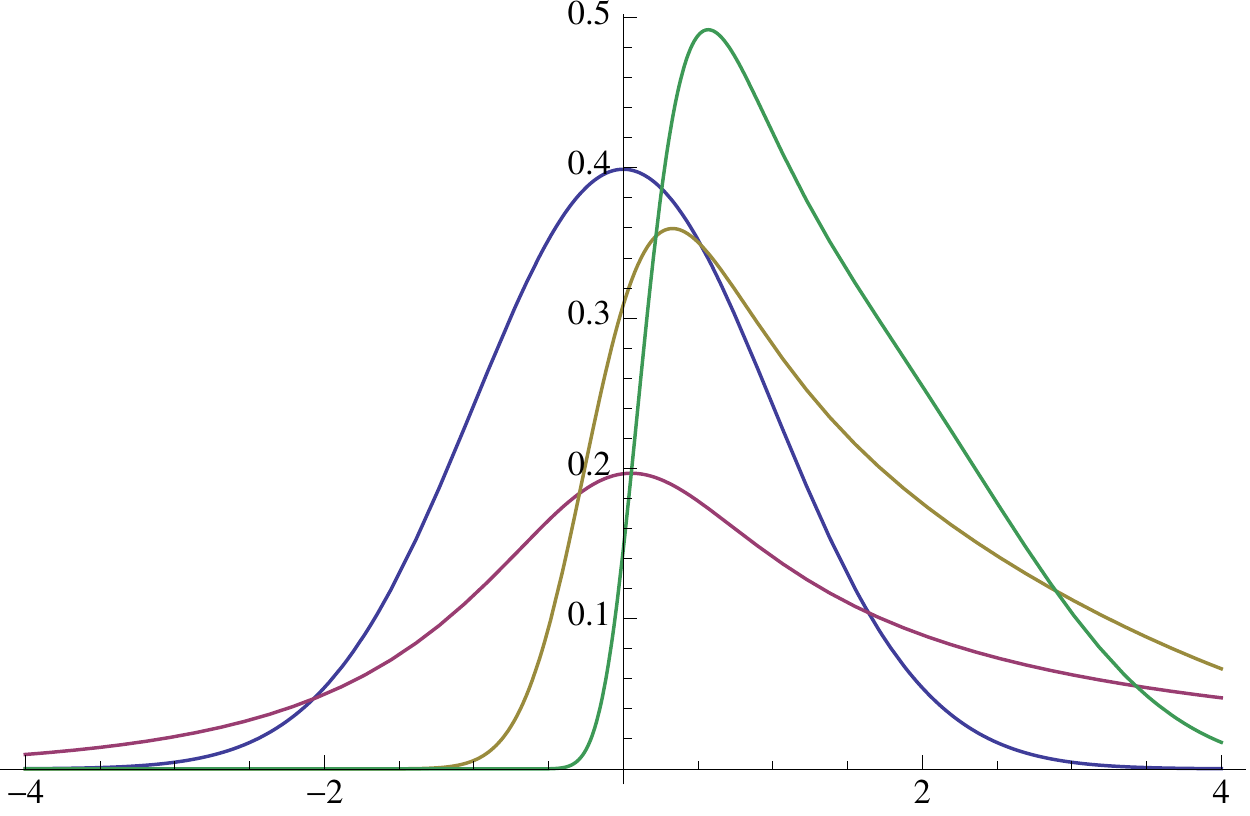}}%
\caption{Plots of the SAS-normal of Jones and Pewsey (2009) with $\mu=0,\sigma=1$ and, on the left picture, $(\delta,\eta)=(0,1)$ (blue), (-0.5,0.5) (red), (-1,0.5) (yellow) and (-1.5,0.5) (green) and, on the right picture, $(\delta,\eta)=(0,1)$ (blue), (-0.5,0.5) (red), (-1,1) (yellow) and (-1.5,1.5) (green). }%
\label{Fig2}
\end{minipage}
\end{center}
\end{figure}

\subsection{Family 3: Two-piece or scale-transformed distributions}

The third family of distributions is, like Family 1, a pure skew family of distributions, the tail-weight being again regulated by applying this construction on e.g. the Student $t$ distribution. While Family 1 can be best understood via the expression ``modulating symmetry'', Family 3 can be described as follows: take a symmetric distribution, change the weight on both sides of the center (re-scaling operation), and glue both parts together! This nice interpretation definitely represents an advantage of these distributions pioneered by Fechner~(1897). It is therefore not astonishing that this intuitive model has been applied by the Bank of England and the Sveriges Riksbank in their probabilistic forecast of future inflation to show that the deviation from the central  forecast might be asymmetric\footnote{According to Wallis~(2014), this application brought the two-piece normal distribution to ''public attention in the late 1990s''.}. In recent years, researchers even went one step further in generality by adding a scaling function to the initial density $f$, hence extending the image from above  by now re-weighting continuously all points on the real line. This is why I opted for the title ``Two-piece or scale-transformed distributions''; it is also for the same reason that Chris Jones in Jones~(2014a) categorizes them as Family 3 and Family 3A.

Under their most general form, univariate two-piece-$f$ distributions possess densities of the form
\begin{equation}\label{2p}
\frac{a(\delta)}{\sigma}\left\{\begin{array}{ll}
f\left(s_\ell(\delta)\frac{(x-\mu)}{\sigma}\right)&\mbox{if}\,x<\mu\\
f\left(s_r(\delta)\frac{(x-\mu)}{\sigma}\right)&\mbox{if}\,x\geq\mu,
\end{array}
\right.
\end{equation}
with $\delta\in\R$ the skewness parameter, $s_\ell(\cdot)$ and $s_r(\cdot)$ left- and right-scaling functions, and  resulting normalizing constant 
$$
a(\delta):=\frac{2}{\frac{1}{s_\ell(\delta)}+\frac{1}{s_r(\delta)}}.
$$ 
The probability mass on the left side of $\mu$ is $s_r(\delta)/(s_\ell(\delta)+s_r(\delta))$, while on the right side it is $s_\ell(\delta)/(s_\ell(\delta)+s_r(\delta))$. This rescaling on both sides is the idea put forward by Fechner~(1897) with $f$ the normal density, and the general   construction~\eqref{2p} has been examined in  Arellano-Valle \emph{et al.}~(2005). The reader is thus referred to that paper for the numerous interesting and good properties of two-piece distributions. One particular choice, suggested in Hansen~(1994) in combination with the Student $t$ and in Mudholkar and Hutson~(2000) with the normal distribution, corresponds to $s_\ell(\delta)=1/(1-\delta)$ and $s_r(\delta)=1/(1+\delta)$ for $\delta\in(-1,1)$; conveniently, $a(\delta)=1$ in that case. Mudholkar and Hutson~(2000) term the resulting two-piece-skew-normal distribution \emph{epsilon-skew-normal} (they use $\epsilon$ as parameter instead of my $\delta$), and investigate its properties. Another famous representative of~\eqref{2p} is the \emph{inverse scale factors (ISF)} model introduced by Fern\'andez and Steel~(1998). As the well-chosen name indicates, $s_\ell(\delta)=1/s_r(\delta)=\delta$ for $\delta>0$, yielding $a(\delta)=2/(\delta+1/\delta)$. Multivariate extensions of the latter choice, built along the same lines as the marginal-wise transformations   in Family~2, are studied in Bauwens and Laurent~(2005) and Ferreira and Steel~(2007a), while Chapter 3 of the thesis of Delphine Cassart from the Universit\'e libre de Bruxelles (Cassart~2007) does the same with the former (epsilon) choice. As for the other flexible families, an abundance of papers besides those mentioned above also propose parametric two-piece distributions, such as skew-Laplace or skew-exponential power distributions; see Wallis~(2014) for references. Figure~\ref{Fig3} contains density plots of  ISF-skew-normal and epsilon-skew-$t$ distributions (the names should by now be self-explanatory). 

\begin{figure}
\begin{center}
\begin{minipage}{140mm}
\resizebox*{14cm}{!}{\includegraphics{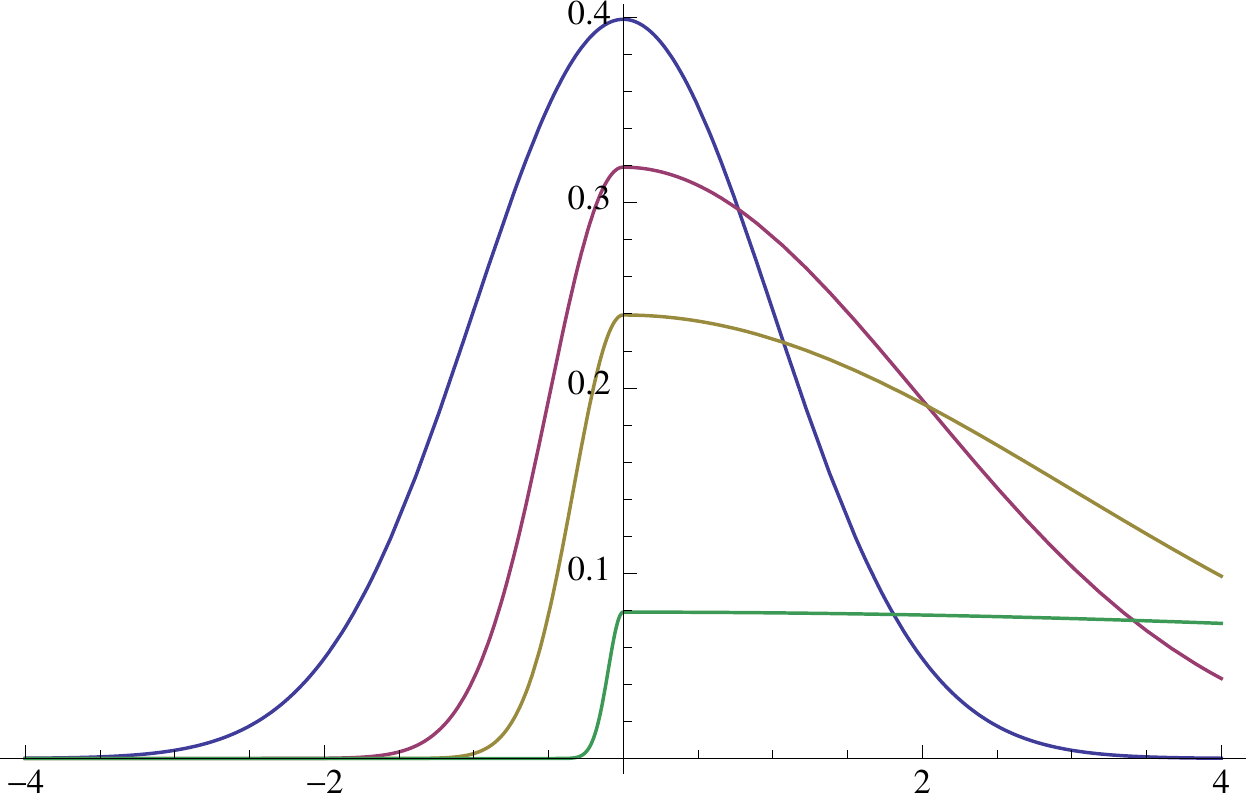}\hspace{1cm}\includegraphics{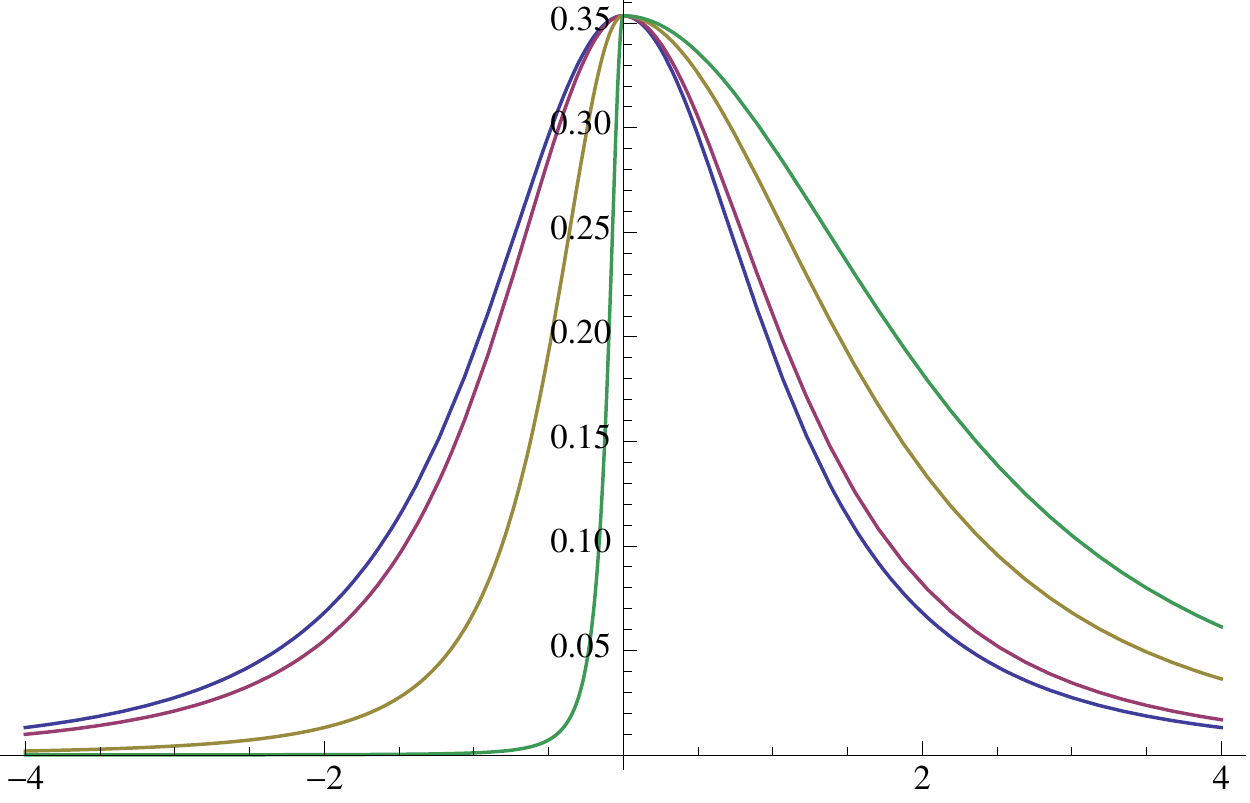}}%
\caption{Plots of the ISF-skew-normal (Fern\'andez and Steel~1998) density with $\mu=0,\sigma=1$ (left) and of the epsilon-skew-$t$ (Hansen~1994) density  with $\mu=0, \sigma=1,\nu=2$ (right),  for $\delta=1$ and 0 (blue), 2 and 0.1 (red), 3 and 0.5 (yellow), and~10 and 0.9 (green). }%
\label{Fig3}
\end{minipage}
\end{center}
\end{figure}

For inferential aspects of two-piece distributions, I refer the reader to Jones and Anaya-Izquierdo~(2011), where it is shown that two-piece distributions actually enjoy strong parameter orthogonality compared to other flexible four-parameter models,  Cassart \emph{et al.}~(2008) for a Le Cam approach and the various cited papers involving Mark Steel for insightful Bayesian aspects.

Scale-transformed distributions, the ``younger brothers of two-piece distributions'', are actually strongly linked to Family 2, as can be read from their densities
\begin{equation}\label{scale}
2f(H^{-1}(x)),
\end{equation}
where the mapping $H:\R\rightarrow\R$ satisfies $H(x)-H(-x)=x$ (for simplicity I have exchanged here the roles of $H$ and $H^{-1}$ compared to Family~2). This strongly resembles \eqref{LP10} but without a Jacobian; the condition on $H$ is the cause thereof. Note that the latter condition, when differentiated, yields $H'(x)+H'(-x)=1$, which is exactly the requirement on skewing functions in Family 1. This again is no coincidence: Families 1-3 are linked, as nicely explained in Jones~(2014a). Densities of the form~\eqref{scale} have been investigated and brought to the statistical community in Jones~(2010) and, mainly, Jones~(2014b) and Fujisawa and Abe~(2013); see those papers for their properties and for further references, especially regarding the classical Cauchy-Schl\"omilch transformation underpinning the entire construction. 
Finally, multivariate extensions of~\eqref{scale} are analysed in Fujisawa and Abe~(2014), following again the independent-component-approach described for Family 2, and Jones~(2014c) via marginal transformations from multivariate skew-symmetric distributions of Family 1.

\section{The future of flexible modelling: open questions and challenges}\label{sec:fut}

The number of existing proposals for flexible distributions has become  huge over the past decades and does not cease to increase. Papers proposing new skew and heavy-/light-tailed distributions  have become common currency in the statistical literature, which is both good and bad. Good because each article does  contribute to the repository of knowledge on statistical modelling, some with very strong  impact on the community, and bad because it gets more and more complicated for the practitioner to choose the ``best'' model among this zoo of new proposals. This is why, besides continuing to improve on existing models, we should also use the current momentum of interest in flexible modelling to heave this research domain on a new level by structuring the collection of distinct models (critical comparison of existing families both on theoretical and practical grounds) and hence make this research stream easier accessible and applicable to people outside the community, be it statisticians, researchers from other disciplines or practitioners. Jones~(2014a) has successfully started this task, and I hope to further contribute to it by addressing more directly what open problems and research questions should be tackled in the future.

I of course by no means claim to be in a position to say which are the main future problems in flexible modelling! However, from several personal reflections, readings of recent papers and discussions with colleagues from the domain (\emph{inter alia} T. Abe, A. Azzalini, A. B\"ucher, H. Dette, Y. Dominicy, M. Hallin, M. C. Jones, A. E. Koudou, D. Paindaveine, A. Pewsey, J. Rubio), I have set up the following list of open questions and challenges (OQC's) that are worth  being addressed in the future.

\

\noindent\underline{OQC1: What desirable properties should a flexible model possess?}\vspace{1mm}

\noindent The history of flexible modelling reveals that the fathers of this research domain had very divergent views on what is a ``good flexible distribution''. For instance, de Helguero's viewpoint  on the necessity of a generating mechanism allowing to explain data generation, his main criticism to Edgeworth's and Pearson's proposals,  has, in turn, been criticized in Pretorius (1930) with the following words: ``The superiority of one frequency function over another depends rather on the success with which that function can be applied to graduate data than on the manner in which it originated''; see Section 1.4 of Johnson (1949) for further details.

Given this discrepancy, it is evident that even today we have divergent opinions on what would be good properties that a flexible model should ideally possess. Nevertheless, I undertake the endeavor of writing out what I consider as desirable structural properties  (I will not consider fitting issues, see OQC2 for that matter).
\begin{itemize}
\item \emph{finite number of well interpretable parameters:} optimally, one skewness and one tail-weight parameter are added to location and scale/scatter (for the sake of clarity: the term ``one'' does not stand for the dimension of the parameter). Evidently, in mixture models one also has to take into account the mixing parameter. A larger number of parameters entails the risk of a loss in interpretability. 
\item \emph{well-separated roles for skewness and tail-weight parameters:} Measures of skewness such as those proposed in Arnold and Groeneveld~(1995) and Critchley and Jones~(2008) should ideally not depend on the tail-weight parameter. Two-piece constructions are particularly viable in this respect. Similarly, kurtosis measures, based on quantiles (e.g., Balanda and MacGillivray~1988) or on moments (e.g., Ley and Paindaveine~2010, Theorem 3.2), ought not be effected by the skewness parameter; here the SAS-transformation constitutes a good choice. For further information on this topic, see Rubio \emph{et al.}~(2013) (where the van Zwet ordering is also discussed) and Jones \emph{et al.}~(2013), respectively.
\item \emph{good parameter estimation and tractability:} once a good-fitting model is determined, it is of utmost importance to be able to correctly estimate the diverse parameters and then to produce calculations with that model, e.g. to predict the risk of exceeding a certain threshold value. 
\item \emph{clear stochastic properties:} this point is related to clear roles for each parameter and good tractability. It is desirable for any distribution to display well-identified conditions as to when, for example, a distribution is uni- or bi-modal. In the multivariate case, properties like closure under marginalization  seem also desirable.
\item \emph{varying shapes:} optimally, a flexible model is able to exhibit as many distinct shapes as possible. For instance, we may wish that a distribution  covers (nearly) all values of AG-skewness or of a given kurtosis measure.
\item \emph{good inferential properties:} besides good parameter estimation, a flexible model should not be prone to inferential problems prohibiting its use as, for instance, alternative in tests for normality (see OQC6). This explains the numerous attempts to remove the singularities linked to Family 1. Parameter orthogonality is, in this respect, an essential attribute of a flexible model (see Jones and Anaya-Izquierdo~2011 or Ley and Paindaveine~2014). 
\item \emph{data generating mechanism:} although in the discussion Ley and Paindaveine~(2014) I was still reluctant towards the usefulness of this criterion, I have in the preparation of the present paper changed my mind and now also consider it as an advantage to be able to explain what causes might have led to a certain behavior of the data and not to confine my-self to just finding the best fitting model. The delicateness of the question is well underlined by the Pretorius citation above. 
\end{itemize}

I recommend the reader to also go through the criteria of Jones~(2014a) and the related discussions. Although formulated under a different form, there is a strong overlapping between my and Chris Jones' criteria, indicating that these are indeed important requirements (their validity is further underlined by the opinions of the various discussants of Jones~2014a). Anticipating on OQC4, good new flexible distributions should thus fulfill (most of) these desiderata. In line with the present OQC, I also recommend the reading of Ferreira and Steel~(2007b).\\

\noindent\underline{OQC2: The curse of abundance: which distribution to be used when?}\vspace{1mm}

\noindent Besides the theoretical comparison of flexible families, another crucial issue consists in providing hints on what family ought to be used with what type of data. Of course, this question is extremely difficult to answer, and I will try to formulate diverse ways of proceeding.
\begin{itemize}
\item The classical procedure would consist in choosing a set of flexible models, say, the skew-normal, skew-$t$, two-piece-$t$ and SAS-logistic, and then use model selection criteria like likelihood maximization, the Akaike information criterion or the Bayesian information criterion. The best model will readily be obtained, sometimes at the cost of complex numerical maximizations. However, if the initial sub-selection is performed in an arbitrary way, then we may only optimize our fitting among distributions that present a low degree of fitting compared to other models. Moreover...
\item ... the interesting recent paper Charemza \emph{et al.}~(2013) has shown that distinct skew-normal distributions may be undistinguishable in certain circumstances!  Their solution  in such a case, from the practitioner's perspective, is to select the model on grounds of parameter interpretation, stressing the importance of this requirement in my listing in OQC1.
\item Some problems very naturally guide us towards the flexible model from which the data have (probably) been generated. Coming back to the examples of the Introduction, it is clear by construction that the error term in Example~\ref{exo1} stems from Family 1 (this is an immediate exercise for the interested reader), the internet traffic data from Example~\ref{exo2} can be well-described by members of Family 2 which allows to control left and right tail-weight, and the BMI data of Example~\ref{exo3}, given their bi-modal structure, seem best fitted by mixture models. Advice of experts from the domains the data originate is evidently helpful.
\item In our discussion Ley and Paindaveine~(2014), we have added  a further criterion to the list of Jones~(2014a) and we term it ``Testability:  natural or satisfactory goodness-of-fit tests for the considered family can be defined''. If satisfied, this criterion (which really means testing the validity of a given family, not testing for the best distribution within a given family!) would of course help to answer OQC2. In our discussion, we have shown that two-piece distributions actually do satisfy the criterion and provided a testing strategy based on the idea of de-scaling both half-distributions. In other families, the situation is more delicate. In Family 1, one can resort to the property that any even function of a skew-symmetric random vector annihilates the effect of the skewing function, leaving only the symmetric part; working with evenly transformed data, the symmetric part can thus be estimated, and the skewing function shall then be detected in a subsequent step within a collection of choices (it appears that, except for the bad matchings identified in Hallin and Ley~2012, the choice of skewing function does not so much influence the final shape of the distribution, see Umbach 2007). Finally, in Family 2, I would suggest transforming the data with several choices of $H_{\deltab,\etab}$,  apply some symmetry or normality test and retain that transformation with the highest p-value (this shall work especially well if our aim is to detect the skewness-inducing transformation). Of course, such procedures are again subject to the criticism mentioned at the first point of this OQC and, contrarily to Family 3, I do not provide a Family-membership-test; my ideas  are merely first steps into that direction. Another natural option would be to check membership by means of kernel density estimation under a specified-family form versus no specified form.
\end{itemize}

Even if OQC2 appears very tough, and near-impossible given the diversity of data sets even within a same domain, we should nevertheless aim to provide a list of recipes for which distribution to use in which circumstance. Such hints would definitely present an important contribution to flexible modelling and its use in other disciplines, especially in finance (what is, in the end, the best skew-heavy-tailed distribution to be combined with GARCH models?), economics and environmental sciences.\\

\noindent\underline{OQC3: Improved multivariate modelling via flexible copulas with flexible marginals?}\vspace{1mm}

\noindent The main appeal of copulas, and the reason for their enormous success story over the past years, lies in the fact that they model separately the dependence structure of multivariate data and the behavior of the individual marginals via the celebrated Sklar theorem of Sklar~(1959), which renders them very attractive, especially in financial areas such as risk management and asset pricing (see, e.g., Cherubini \emph{et al.} 2004). The success of copulas has been further stimulated by the introduction of vine copulas which are tailored for different forms of dependence between the variables (see Kurowicka and Joe 2010). Thus, it is completely appealing to aim at a stronger combination of both research domains\footnote{The 2013 conference ``Non-Gaussian Multivariate Statistical Models and their Applications'' precisely followed this goal.}: \emph{flexible copulas} for the dependence structure, and flexible univariate distributions for the margins. This  clearly generalizes  the dependence structure described for certain members of Family 2 and 3, where instead of a full copula only a correlation matrix is added to independent component models.  A neat treatment of OQC3 will most certainly result in promising constructions and should have a strong impact also outside statistics, especially in finance. For further recent information about copulas, I refer to the special volume 154(1) of the Journal de la Soci\'et\'e Fran\c caise de Statistique dedicated to this topic.\\

\noindent\underline{OQC4: Holy grail question: does THE flexible distribution exist?}\vspace{1mm}

\noindent Very naturally, each person has his/her personal favorite  flexible model, and divergent viewpoints on the subject have helped shape this research landscape (recall the stormy beginnings of flexible modelling!). Such heated debates, of course under a less aggressive form, would certainly also nowadays push forward the quest for the ultimate flexible model that could be used for ``all'' purposes and would satisfy ``all'' requirements one has on a flexible distribution. Can such a universal distribution exist, or is it at least possible to come close to it? And if so, how could this discovery be achieved?

One potential solution might be to efficiently combine distinct flexible proposals. Rubio \emph{et al.}~(2013) combine the symmetric SAS transformation with (i) the two-piece construction and (ii) Azzalini-type skewing functions, while Steel and Rubio~(2014) propose merging Ferreira-Steel (2006) with Ley-Paindaveine~(2010); further investigations in this direction, especially under the form of mixture distributions with flexible components, should allow to go beyond the state-of-the-art.\\

\noindent\underline{OQC5: Flexible modelling on other supports?}\vspace{1mm}

\noindent Besides the multivariate setting, a further challenge consists in consolidating respectively developing flexible distributions on  supports other than $\R^k$; the word ``consolidation'' is meant for positive data and data on the unit circle,  ``developing'' for data on the unit hypersphere $\mathcal{S}^{k-1}:=\{\xb\in\R^k:\xb'\xb=1\}, k>2$. Indeed, somehow paradoxically, while circular ($k=2$) flexible models have recently met a strong success under the impetus of researchers like Abe, Jones, Kato and Pewsey, the same cannot be said when $k>2$ where, to the best of my knowledge, so far only mixture models have been proposed. Palliating this need is an important open problem.\\

%

\noindent\underline{OQC6: Improved tests for normality and symmetry via flexible models?}\vspace{1mm}

\noindent Testing for normality and symmetry is a classical topic in statistics. Famous normality tests are the Jarque-Bera, Shapiro-Wilks and Lilliefors tests, univariate symmetry is recurrently tested by means of Kolmogorov-Smirnov- and Cram\'er-von Mises-type tests; interesting alternative symmetry tests are the triples test of Randles \emph{et al.}~(1980) or the runs test of McWilliams~(1990). For tests of multivariate symmetry under its different forms, see Serfling~(2006).

It is clear that flexible models, besides their role of ``improved description of reality'', should also serve the purpose of insightful alternatives to the null hypotheses of normality and symmetry. Their specific forms allow to test for varying departures from normality (most naturally under a likelihood ratio form), be it in terms of skewness or kurtosis. Razali and Wah (2011)  provide a broad power comparison of several classical normality tests, while Jones and Pewsey~(2009)  compare the performance of SAS-based  normality tests to classical competitors. It would be of great interest  to build likelihood ratio tests for normality within each flexible family, and then compare the overall performance of the resulting tests. While it is obvious that the Family-$\ell$-test is the best under  Family $\ell$, it is far from clear if one test is able to exhibit good power under general alternatives. If such a test can be found, it is highly probable that this test can become a competitor for the above-mentioned classical tests.

The same holds true when testing for symmetry. Tests that behave optimally against a given family have been studied, e.g., for two-piece distributions in Cassart \emph{et al.}~(2008) and for Ferreira-Steel type distributions in Ley and Paindaveine~(2009). Extending this effort to other flexible families and comparing the ensuing testing procedures in terms of their power seems a promising research topic.

%

\

\

Arrived at this point, each reader will agree and disagree with me on the relevance of certain OQC's (and certainly have further OQC's in mind); some may be willing to take up a given challenge, others may (hopefully) benefit from certain ideas for their own research. The subdivision into underlined OQC's, which are sometimes provocatively formulated, follows a very precise goal: stimulate reflections on how we may/shall address flexible modelling in the future and open up discussions on this thrilling and ageless research direction!

\

\bigskip
\noindent\textbf{Acknowledgements}
\vspace{4mm}

I would like to thank the Soci\'et\'e Fran\c caise de Statistique for awarding me, jointly with Aur\'elie Fischer, the Marie-Jeanne Laurent-Duhamel Prize 2014. 
I also thank the Fonds National de la Recherche Scientifique, Communaut\'{e} fran\c{c}aise de Belgique, for support via a Mandat de Charg\'{e} de Recherche FNRS.

\end{document}